\documentstyle[12pt]{article}
\newtheorem{theorem}{Theorem}
\newtheorem{proposition}[theorem]{Proposition}

\newtheorem{corollary}[theorem]{Corollary}

\def\R{{\bf R}} 
 
\def\N{{\bf N}}
\def\T{\bf T}  
\def\C{{\bf C}}
\def\b{\beta}
\def\t{\theta}
\def\la{\langle}
\def\be{\begin{equation}}
\def\ee{\end{equation}}
\def\ra{\rangle}
\def\ds{\displaystyle}

\def\im{{\rm Im}}
\def\RS{Ray\-leigh-Schr\"o\-din\-ger}
\def\Sc{Schr\"odinger}
\def\PT{{\cal P}{\cal T}}
\def\P{{\cal P}}
\def\T{{\cal T}}
\def\l{{\lambda}}
\def\r{\rho}
\def\S{{\cal S}}
\def\op{o\-pe\-ra\-tor}
\def\arg{{\rm arg}}
\def\Im{{\rm Im}}
\def\Cinf{C_0^\infty(\R^d)}
\date{}
\begin{document}
\baselineskip=21pt
\title{Canonical Expansion of $\PT-$Symmetric Operators and Perturbation 
Theory}
\author{E.Caliceti\footnote{e-mail: caliceti@dm.unibo.it}  \\ Dipartimento 
di 
Matematica, Universit\`{a} di Bologna
\\40127 Bologna, Italy
\\ 
S.Graffi\footnote{On leave from Dipartimento di Matematica, Universit\`{a} 
di 
Bologna, Italy; e-mail: graffi@mathcs.emory.edu. graffi@dm.unibo.it}
 \\ Department of Mathematics and Computer Science\\ Emory University, 
Atlanta, Ga 30322. U.S.A.}
\maketitle 
\vskip 12pt\noindent 
\begin{abstract}
 { \noindent\baselineskip =16pt 
Let $H$ be any $\PT$ symmetric Schr\"odinger
operator of the type 
$\;-\hbar^2\Delta+(x_1^2+\ldots+x_d^2)+igW(x_1,\ldots,x_d)$
 on $L^2(\R^d)$, where $W$ is any odd homogeneous polynomial and
$g\in\R$.  It is proved that $\P H$ is
self-adjoint and that its eigenvalues coincide (up to a sign) with the
singular values of $H$, i.e. the eigenvalues of
$\sqrt{H^\ast H}$. Moreover we explicitly construct the canonical 
expansion of $H$ 
and determine the
singular values $\mu_j$ of $H$ through the Borel summability  of their divergent 
perturbation theory. The singular values yield  estimates of the location of the 
eigenvalues $\l_j$ of $H$ by Weyl's inequalities.
}
\end{abstract}
\vskip 12pt\noindent 
\section{Introduction and statement of the results}
\setcounter{equation}{0}%
\setcounter{theorem}{0}%
 
A   Schr\"odinger operator $H=-\Delta+V$ acting on ${\cal H}=L^2(\R^d)$ is
called
$\PT-$symmetric if it is left invariant by the $\PT$ operation. While 
generally
speaking $\P$ could be the parity operator with respect to at least one 
variable,
here for the sake of simplicity we consider only the case in which $\P$
is the parity operator with respect to all variables,
$(\P u)(x_1,\ldots,x_d)=u(-x_1,\ldots,-x_d)$, and
$\T$ the complex conjugation (equivalent to time-reversal symmetry)
$(\T u)(x_1,\ldots,x_d):=\overline{u}(x_1,\ldots,x_d)$. The condition
\be
\label{PT}
\overline{V}(-x_1,\ldots,-x_d)=V(x_1,\ldots,x_d)
\ee
defines the $\PT-$symmetry on the potential $V(x_1,\ldots,x_d)$. The
$\PT$-symmetric  operators are currently the object of  intense 
investigation
because, while not self-adjoint, they admit in many
circumstances a real spectrum. Hence the investigation is motivated (at
least partially), by an attempt to remove the self-adjointess condition on
the observables  of standard quantum mechanics (see
e.g.\cite{Ah},\cite{Be1},\cite{Be2},\cite{Be3},\cite{Cn1},\cite{Cn2},
\cite{Cn3},
\cite{Zn1},\cite{Zn2}).

The simplest and most studied class of $\PT$ symmetric operators is 
represented
by the {\it odd anharmonic oscillators with purely imaginary coupling} in
dimension one, namely the maximal differential operators in $L^2(\R)$
\be
\label{od}
Hu(x) :=[-\frac{d^2}{dx^2} +x^2+igx^{2m+1}], \quad g\in\R,\quad 
m=1,2,\ldots
\ee
  It has
long been  conjectured (Bessis Zinn-Justin), and recently proved
\cite{Shin},
\cite{Tateo}, that the spectrum
$\sigma(H)$ is real for all $g$;  there are however examples of
one-dimensional
$\PT$-symmetric operators with {\it complex} eigenvalues\cite{Cn1}. 

Now recall that there is a natural additional notion of  spectrum
associated with a non-normal operator $T$ in a Hilbert space which is by
construction real.  Any closed operator
$T$ admits a {\it polar decomposition} (\cite{Ka}, Chapt. VI.7) $T=U|T|$,
where $|T|$ is self-adjoint and $U$ is unitary.  The modulus of $T$ is
the  self-adjoint operator
$\ds |T|=\sqrt{T^\ast T}$. The (obviously real and positive)
eigenvalues of $|T|$ are called the  {\it singular values} of
$T$. In this paper we consider the self-adjoint operator
$\sqrt{H^{\ast}H}$; its eigenvalues $\mu_j;\; j=0,1,\ldots$, necessarily real and 
positive, 
 are the by definition the {\it singular values} of $H$.
A  first immediate question arising in this context is to determine how
these singular values are related to the 
$\PT$-symmetry of  $H$.  A related question is the explicit
construction of the canonical expansion of $H$ (see e.g.\cite{Ka}) in
terms of the spectral analysis of $\sqrt{H^{\ast}H}$, which entails the
diagonalization of $H$ with respect to a pair of  dual
bases (which  do not form a biorthogonal pair);  a further one is
the actual computation of the singular values.   
The determination of the 
singular values reflects directly on the object of physical interest, namely the 
eigenvalues $\l_j;\; j=0,1,\ldots$ of $H$.  If the eigenvalues and the singular 
values are ordered according to increasing modulus, the Weyl 
inequalities (see e.g. \cite{Ho}) indeed yield
\begin{eqnarray}
\label{Weyl1}
\sum_{j=1}^k |\l_j|\leq \sum_{j=1}^k\mu_j,\quad 
|\l_1\cdots\l_k|\leq \mu_1\cdots\mu_k, \quad k=1,2,\ldots
\end{eqnarray}

We intend in this paper to give a
reply   to these questions for the most general class of odd anharmonic
oscillator in $\R^d$. Namely, we consider in $L^2(\R^d)$ the  \Sc\ operator
family
\be
\label{odM}
H(g)u(x):= H_0u(x)+igW(x)u(x),\quad x=(x_1,\ldots,x_d)\in\R^d
\ee
Here:
\begin{enumerate}
\item $W$ is a real homogenous polynomial of odd order $2K+1$, 
$K=1,2\ldots$;
$$
W(\l x)=\l^{2K+1}W(x)
$$
\item  $H_0$ is the \Sc\ operator of the harmonic oscillator in $\R^d$:
\be
\label{HO}
H_0u(x)=-\Delta u(x)+x^2u(x),\quad x^2:=x_1^2+\ldots+x_d^2
\ee
\end{enumerate}
Under these conditions the operator family $H(g)$, which is obviously $\P
T$-symmetric (see below for the mathematical definition), but non
self-adjoint, enjoys the following properties
(proved in \cite{CGM} for $d=1$ and in \cite{Na} for $d>1$; see below for
a more detailed statement):
\begin{enumerate}
\item The operator
 $H(g)$, defined as the closure of the minimal differential operator 
$\dot{H}(g)u=-\Delta u(x)+x^2u(x)+igW(x)$, 
$u\in C_0^\infty(\R^d)$, generates a holomorphic
operator family with compact resolvents with respect to 
$g$ in some domain ${\cal S}\subset\C$,  with
$H(g)^\ast=H(\overline{g})$.  An 
operator family $T(g)$ depending on the complex variable $g\in \Omega$,
where $\Omega\subset\C$ is open is holomorphic (see
\cite{Ka}, VII.1)  if the scalar products $\langle u,T(g)v\rangle$ are
holomorphic functions of $g\in \Omega$ $\forall\,(u,v)\in T(g)$ and the
resolvent $\ds [T(g)-zI]^{-1}$ exist for at least one $g\in\Omega$. 
\item All eigenvalues of $H_0:=H(0)$ are stable with respect to the
operator family $H(g)$. 
This means (see e.g.\cite{Ka}, VIII.1) that if $\lambda_0$ is
any eigenvalue of $H(0)$ of multiplicity $m$, there is $B(\lambda_0)>0$ 
such
that $H(g)$ has exactly $m$ (repeated) eigenvalues $\lambda_j(g):
j=1,\ldots,m$ near $\lambda_0$ for $g\in{\cal S}$, $|g|<B(\lambda_0)$, 
and $\ds \lim_{g\to 0, g\in{\cal S}}\lambda_j(g)=\l_0$. 
\item The (\RS) perturbation series of each eigenvalue $\l(g)$ of $H(g)$ is
Borel summable to $\l(g)$. 
\end{enumerate}

We denote  $\mu_j(g): j=0,1,\ldots$ the singular values of $H(g): g\in\R$,
i.e. the eigenvalues 
 of $\sqrt{H(g)^\ast H(g)}= \sqrt{H(-g)
H(g)}$.   
\par\noindent
Our first result concerns the identification of the singular values 
as the eigenvalues of a self-adjoint operator directly associated with
$H(g)$ by the operator-theoretic implementation of the recently isolated 
{\it
pseudohermiticity} notion (\cite{Ah},\cite{KS},\cite{Mo},\cite{We}) in
terms of the $\P$ symmetry itself. 

Consider indeed  the operator family
$Q(g):=\P H(g)$. We will show that $D(Q(g))=
D(H(g))$. Since $[\P,H_0]=0$, the explicit action of $Q(g)$ is
\begin{eqnarray*}
Q(g)u(x)&=&H_0u(-x)+
 ig{W}(-x)
u(-x)
\\
&=&H_0u(-x)-igW(x)u(-x) =H(-g)\P u(x)
\end{eqnarray*}
Then we have:
\begin{theorem}
\label{teor1}
Let $Q(g)$ be defined as above and $
Q^\prime(g):=H(g)\P$. Then:
\begin{enumerate}
\item
If $g\in\R$ the operator families $ Q(g)$ and $ Q^\prime(g)$ are
self-adjoint.
\item The operator
family $Q(g)$ defined on $D(H(g))$ is 
holomorphic  with compact resolvents at least for $g$ 
in a neighbourhood of $\R_+$.
\item If $g\in\R$ the eigenvalues of $Q(g)$ and of $\sqrt{H(g)^\ast H(g)}$
coincide (up to the sign);
\end{enumerate}
\end{theorem} 
{\bf Remarks}
\begin{enumerate}
\item
$H(g)^\ast=H(-g)$ for $g\in\R$ by $\PT$-symmetry. Hence the relation 
$Q(g)=\P H(g)=H(-g)\P=H(g)^\ast\P$ can be equivalently written $\ds \P
H(g)\P^{-1}=H(g)^\ast$ which is the $\P-$ pseudohermiticity
property of $H(g)$ \cite{Mo}.  
\item
The eigenvalues $\mu$ of
the operator
$Q(g)$ clearly solve the generalized spectral problem $H(g)u=\mu \P u$
(for this notion, see \cite{Ka}, \S VII.6). Explicitly:
\be
\label{sv}
(H_0 +ig W)u(x)=\mu (\P u)(x)
\ee
By the above theorem the singular values coincide (up to a sign) with the
generalized eigenvalues. 
\end{enumerate}
 As a consequence of this, we obtain the
explicit canonical expansion of $H(g)$ in terms of the eigenvectors
$\psi_k$ of $Q$ and of the $\P$ operation:
\begin{corollary}
\label{can}
\par\noindent
Let $\{\psi_k(g)\}:k=0,1,\ldots$ be the eigenvectors of $Q(g)$, and
$\mu_k$ the corresponding eigenvalues (counting multiplicy). Then
$H(g)$ admits the following canonical expansion
\be
\label{canonical}
H(g)u=\sum_{k=0}^{\infty}\mu_k\langle u,\psi_k\rangle\P\psi_k, \quad u\in
D(H(g))
\ee
\end{corollary}
{\bf Remarks}
\begin{enumerate}
\item
Since $\P\psi_n(x)=\psi_n(-x)$, the canonical expansion (\ref{canonical})
entails that $H(g)$ can be dia\-go\-na\-lized in terms of the (repeated)
real singular values $\mu_n$ and of the pair of orthonormal bases
$\{\psi_n(x)\}$ and $\{\psi_n(-x)\}$.
\item For a general operator with compact resolvent the canonical
expansion reads
\be
\label{cc1}
Tu=\sum_{k=0}^{\infty}\mu_k\langle u,\psi_k\rangle\psi^\prime_k, \quad
u\in D(T)
\ee
Here $\{\mu_k\}$ is the sequence of singular values of $T$, 
$\psi_k$ the corresponding eigenvectors, but the dual basis
$\{\psi^\prime_k\}$ is a priori unknown. In this case it is simply
the $\P$-dual basis $\P\psi_k$. Remark that the orthogonal bases $\psi_k$ 
and
$\P\psi_k$ do
not form a biorthogonal set. 
\item
The expansion (\ref{canonical}) is useful even when all eigenvalues of 
$H(g)$ are
real, because $H(g)$ is not normal and the spectral theorem does not
hold. 
\item
Finally we note the following relation involving nonzero eigenvalues and 
eigenvectors on one side and nonzero singular values and corresponding 
eigenvectors 
on the other side: if $H(g)\psi_k=\mu_k(g)\P\psi_k$, and 
$H(g)\phi_l=\lambda_l(g)\phi_l$, then:
\begin{equation}
\label{relazione}
\lambda_l(g)\la\phi_l,\P\psi_k(g)\ra=\mu_k(g)\la\phi_l,\psi_k(g)\ra
\end{equation}
 One has indeed (omitting the 
$g$-dependence):
$$
\lambda_l\la\phi_l,\P\psi_k\ra=\la H \phi_l,\mu_k^{-1}H\psi_k\ra=
\langle H \phi_l,\mu_k^{-1}H^\ast H\psi_k\ra=\mu_k\la\phi_l,\psi_k\ra
$$
\end{enumerate}
 Our third result deals with the actual computation of the
singular values 
$\mu_j(g)$. To formulate the result,  remark that  the closed subspaces
$\P {\cal H}$ and $(1-\P) {\cal H}$ are
invariant under $H_0$ because $[\P
,H_0]=0$ . The operator $\P H_0$ has the
same eigenvectors of $H_0$, but the eigenvalues $\l_l=2l_1+\ldots+2l_d+d$,
$l_k=0,1,\ldots$, $k=1,\ldots,d$,  of $H_0$ split into {\it even }
and {\it odd} eigenvalues. More precisely,  introduce the usual {\it
principal quantum number} $l:=l_1+\ldots+l_d: l=0,1,\ldots$. Then the
eigenvalues of $H_0$ are $\l_l=2l+d$, with multiplicity $m(l)=l^{d-1}$. The
eigenvalues of $\P H_0$ are 
\begin{equation}
\label{av}
\l_l=\left\{\begin{array}{c} 2l+d,  \quad l\;{\rm even} \\
- (2l+d),  \quad l\;{\rm odd}
\end{array}\right.
\end{equation}
The corresponding eigenvectors will be $\P$ even and $\P$ odd, 
respectively.
We then have:
\begin{theorem}
\par\noindent
\begin{enumerate}
\item
All eigenvalues  $\l_l$ of $\P H_0$ are stable as eigenvalues
$\mu_j(g):j=1,\ldots,m(l)$ of 
$Q(g)$  
as $|g|\to 0$, $g\in {\cal
S}_1\cup {\cal S}_2$ where:
\begin{eqnarray}
\label{S1}
{\cal S}_1&:=&\{g\in\C\setminus\{0\}: -\frac{\pi}{2}<{\rm arg}g 
<\frac{\pi}{2}\}
\\
{\cal S}_2&:=&\{g\in\C\setminus\{0\}: \frac{\pi}{2}<{\rm arg}g 
<\frac{3\pi}{2}\}
\label{S2}
\end{eqnarray}
\item All eigenvalues $\mu_j(g):j=1,\ldots,m(l)$ are holomorphic on the
Riemann surface 
 sector
$$
{\cal S}_{K,\delta}:=\{g\in\C: 0<|g|<B(\delta);
-(2K+1)\frac{\pi}{4}+\delta<{\rm arg}\,(g)<(2K+1)\frac{\pi}{4}-\delta\}
$$
where $\delta>0$ is arbitrary.
\item 
The Rayleigh-\Sc\ perturbation expansion for any eigenvalue
$\mu_j(g):j=1,\ldots,m(l)$ of
$Q(g)$ near the eigenvalue $\l_l$ of $\P H_0$ for $|g|$ small is Borel
summable to  $\mu_j(g):j=1,\ldots,m(l)$.
\end{enumerate}
\end{theorem}
{\bf Remark}
\par\noindent
Let $\mu(g)$ be a singular value near an unperturbed eigenvalue $\l$. The Borel 
summability (see e.g.\cite{RS}, Chapter XII.5) means that it can be uniquely 
reconstructed 
through its divergent perturbation expansion $\ds 
\sum_{s=0}^\infty\mu_sg^s,\;\mu_o=\l$ in the following way:
\be
\label{Borel}
\mu(g)=\frac1{q}\int_0^\infty\mu_B(gt)e^{-t^{1/q}}t^{-1+1/q}\,dt
\ee
Here $\ds q=\frac{2K-1}{2}$ and $\mu_B(g)$, the {\it Borel transform of order $q$} 
of the perturbation series, is defined 
by the power series
$$
\mu_B(g)=\sum_{s=0}^\infty\frac{\mu_s}{\Gamma[q(s+1)]
}g^s
$$ 
which has a positive radius of convergence. The proof of (\ref{Borel}) consists 
precisely in showing that $\mu_B(g)$ has analytic continuation along the real 
positive axis and that the integral converges for some $0\leq g<B$, $B>0$.
\par\noindent
{\bf Example} 
\par\noindent
The  H\'enon-Heiles potential, i.e. the third degree polynomial
in $\R^2$
$$
W(x)=x_1^{2} x_2
$$
 
\section{Proof of the results}
\setcounter{equation}{0}%
\setcounter{theorem}{0}%
 
Let us begin by a more detailed quotation of Theorem 1.1 of \cite{Na}. The
results are more  conveniently  formulated in the variable $\beta=ig$
instead of
$g$. 

Let $\b\in\C$, $0<|{\rm arg}\,\b|<\pi$, and let $\dot{H}(\b)$ denote the
minimal differential operator in $L^2(\R^d)$ defined by $-\Delta+x^2+\b
W(x)$ on $C_0^\infty(\R^d)$, with $x^2=x_1^d+\ldots+x_d^2$. Then
\begin{itemize}
\item [(N1)] $\dot{H}(\b)$ is closable. Denote 
 $H(\beta)$ its closure. 
\item [(N2)] $H(\beta)$  represents a pair of type-A holomorphic
families in the sense of Kato for
$\ds 0<{\rm arg }\b<{\pi}$
and $\ds -\pi<{\rm arg
}\b<0$, respectively, with $H(\b)^\ast=H(\overline{\b})$.  Recall that
an operator family $T(g)$ depending on the complex variable $g$
belonging to some open set $\Omega\subset\C$ is called type-A holomorphic
if its domain $D$ does not depend on $g$ and the scalar products $\langle
u,T(g)\rangle$ are holomorphic functions for $g\in D$ $\forall\;(u,v)\in
D$. A general theorem of Kato (\cite{Ka}, VII.2) states that the isolated
eigenvalues of a type-A holomorphic family are locally holomorphic
functions of $g\in D$ with at most algebraic branch points. 
\item [(N3)] $H(\beta)$ has compact resolvent $\forall\,\b\in\C$, $0<|{\rm
arg}\,\b|<\pi$. 
\item[(N4)]
All eigenvalues of 
$H_0=H(0)$  are stable with respect to the operator family $H(\b)$ for
$\b\to 0$,  $0<|{\rm arg}\,\b|<\pi$;
\item[(N5)] Let $\b\in\C$, $\sigma\in\C$, $0<|{\rm arg}\,\b|<\pi$, 
$-\pi+{\rm arg}\,\b \leq {\rm arg}\,\sigma \leq {\rm arg}\,\b$, and let
$\dot{H}_\sigma(\b)$ denote the minimal differential operator in 
$L^2(\R^d)$
defined by
$-\Delta+\sigma x^2+\b W(x)$ on $C_0^\infty(\R^d)$. Then
$\dot{H}_\sigma(\b)$ is sectorial (and hence closable) because its
numerical range is contained in the half-plane $\{z\in\C: -\pi+{\rm
arg}\,\b \leq {\rm arg}\,\sigma \leq {\rm arg}\,\b\}$;
\item[(N6)] Let ${H}_\sigma(\b)$ denote the closure of
$\dot{H}_\sigma(\b)$. Let $\sigma\in\C, \sigma\notin ]-\infty,0]$. Then
the operator family $\b\mapsto {H}_\sigma(\b)$ is type-A holomorphic with
compact resolvents for $\b\in {\cal C}_\sigma:=\{\b\in\C: 0<{\rm
arg}\,\b-{\rm arg}\,\sigma <\pi\}$. Moreover if $\b\in\C, {\rm Im}\b >0$,
the operator family $\sigma\mapsto {H}_\sigma(\b)$ is type-A holomorphic 
with
compact resolvents in the half-plane ${\cal D}_\beta=
\{\sigma\in\C: 0<{\rm
arg}\,\b-{\rm arg}\,\sigma <\pi\}$
\end{itemize}
Let us now introduce the operator
\be
\label{dilat}
H(\b,\t)=e^{-2\t}\Delta+e^{2\t} x^2+\b e^{2(K+1)\t}W(x):=e^{-2\t}K(\b,\t)
\ee
For $\t\in\R$ $H(\b,\t)$ is unitarily equivalent to $H(\b)$, ${\rm Im}\b>
0$, via the dilation operator defined by
\be
\label{dilat1}
(U(\t)\psi)(x)=e^{d\t/2}\psi(e^\t x), \qquad \forall\,\psi\in L^2(\R^d)
\ee 
 As a consequence of (N6)
 (see again
\cite{Na}, or also
\cite{Ca}, where all details are worked out for $d=1$, and where the
reader is referred also for the proof of statement( N8) below) we have: 
\begin{itemize}
\item[(N7)] $H(\b,\t)$ defined on $D(H(\b))$ represents a type-A
holomorphic family with compact resolvents for $\b$ and $\t$ such that 
$ s={\rm
arg}\b, \; t=\im\t$ are
variable in the parallelogram ${\cal  R}$ defined as
\be
\label{par}
{\cal  R}=\{(s,t)\in\R^2:0<(2K-1)t+s<\pi, 0<(2K+3)t+s<\pi\}, \;
\ee
Moreover $C_0^\infty$ is a core of $H(\b,\t)$.  The spectrum of $H(\b,\t)$
does not depend on $\t$.  Note that
$(s,t)\in {\cal  R}$ entails that the maximal range of $\b$ is 
$-(2K-1)\pi/4
<\arg\b<(2K+3)\pi/4$ and that the maximal range of $\t$ is $-\pi/4 <\Im\t
<\pi/4$;
\item[N8)] Let $\b$ and $\t$ be such that $(s,t)\in{\cal  R}$. Then:
\begin{itemize}
\item[(i)] If $\l\notin\sigma (K(0,\t))$, then $\l\in\tilde\Delta$, where:
\begin{eqnarray}
\label{unif}
\tilde\Delta:=\{z\in\C:z\notin \sigma(K(\b,\t));
\|[z-(K(\b,\t)]^{-1}\|
\\
\mathrm{is}\; \mathrm{uniformly}\;\mathrm {bounded} 
\;\mathrm {for}\;|\b|\;
\to 0 \} ;
\nonumber
\end{eqnarray}
\item[(ii)] If $\l\in\sigma (K(0,\t))$, then $\l$ is stable with respect to
the operator family $K(\b,\t)$.
\end{itemize}
(N7) and (N8) entail:
\item[(N9)]
Let $\b\in\C$ with $\ds 0<{\rm arg}(\b)<{\pi}$. Then for
any $\delta >0$ and any eigenvalue $\l(g)$ of  $H(\b)$ there exists
$\rho >0$ such that the function $\l(\b)$, a priori holomorphic for
$0<|g|<\rho$, $\ds \delta<{\rm
arg}(\b)<{\pi}-\delta$, 
has an analytic continuation to the Riemann surface sector $\ds
\tilde{\S}_{K,\delta}:=
\{\b\in\C: 0<|\b|<\rho; -(2K-1)\frac{\pi}{4}+\delta<{\rm
arg}(\b)<(2K+3)\frac{\pi}{4}-\delta\}$.
\end{itemize}
{\bf Remarks}
\begin{enumerate}
\item
The stability statement means the following: if $r>0$ is sufficiently 
small, so that
the
only eigenvalue of $K(0,\t)$ enclosed in $\Gamma_r:=\{z\in\C: |z-\l|=r\}$
is $\l$, then there is $B>0$ such that for $|\b|<B$ ${\rm dim}P(\b,\t)=
{\rm dim}P(0,\t)$, where
\be
\label{pro}
P(\b,\t)=\frac{1}{2\pi i}\oint_{\Gamma_r}[z-(K(\b,\t)]^{-1}\,dz
\ee
is the spectral projection of $K(\b,\t)$ corresponding to the points of
the spectrum enclosed in $\Gamma_r\subset\C\setminus \sigma(K(\b,\t))$.
Similarly for $P(0,\t)$.
\item
Starting from the operator $H(\b)$, $\Im\b<0$, analogous results hold for
the operator family $H(\b,\t)$ where this time $\b$ and $\t$ are such that
$s={\rm
arg}\b$ and $t=\im\t$ describe the parallelogram
\be
\label{par1}
{\cal  R}^1=\{(s,t)\in\R^2:-\pi<(2K-1)t+s<0, -\pi<(2K+3)t+s<0\}.
\ee
Moreover, $H(\b,\t)^\ast=H(\overline{\b},\overline{\t})$.
\end{enumerate}
We now set $\b=ig$ and with slight abuse of notation the operator
$H(\b)= H(ig)$ will be denoted $H(g)$. 

 Let once again
$\P$ denote the parity operator in
${\cal H}$
$$
\P \psi(x)=\psi(-x), \qquad \forall\,\psi\in {\cal H}
$$
 $P$ is a self-adjoint, unitary involution, i.e. $\P^2 =I$, and 
$\P W(x)=-W(x)\,\forall\,x\in\R^d$. 

To show Theorem 1.1, let us first state and prove the following preliminary
result:
\begin{proposition}
\label{prop1}
Let $\ds {\cal S}_1,\;{\cal S}_2$ be the complex sectors defined by
(\ref{S1},\ref{S2}). Then:
\begin{itemize}
\item[(1)] $D(\P H(g))=D(H(g)\P)=D(H(g)^\ast\P)=D(\P H(g)^\ast)=D(H(g))$ 
for
all 
$g\in{\cal S}_1\cup {\cal S}_2$;
\item[(2)] $\P H(g))=H(-g)\P$ for all 
$g\in{\cal S}_1\cup {\cal S}_2$. In particular, for $g\in\R$, $\P
H(g)=H(g)^\ast\P$ whence $\P H(g)\P=H(g)^\ast$, i.e. $H(g)$ and $H(g)^\ast$
are unitarily equivalent;
\item[(3)] $\overline{\P H(g)\psi}=H(g)\P \overline{\psi}, 
\,\forall\,\psi\in
D(H(g))$, $\forall\,g\in\R$. 
\end{itemize}
\end{proposition}
{\bf Proof}
\par\noindent
(1) Since $H(g)^\ast=H(-\overline{g})$, and $D(H(g))$ is independent of
$g\in{\cal S}_1\cup {\cal S}_2$, it is enough to prove that, for all 
$g\in{\cal S}_1\cup {\cal S}_2$:
$$
(a)\qquad\;\;D(\P H(g))=D(H(g));\qquad\quad
(b)\qquad\;\;D(H(g)\P)=D(H(g))\;\forall\,g\in S_1\cup S_2
$$
(a) follows from $D(\P)={\cal H}$. As for (b) notice that $u\in D(H(g))$
if and only if $\exists \{u_n\}\in C_0^\infty(\R^d)$ such that $u_n\to u$
and $H(g)u_n\to v=H(g)u$. Then $u_n(-x)\in C_0^\infty(\R^d)\to u(-x)$ and
$H(-g)u_n(-x)\to v(-x)$. Thus $\P u=u(-x)\in D(H(-g))=D(H(g))$, i.e. $u\in
D(H(g)\P)$. Conversely, if $u\in
D(H(g)\P)$ then $u(-x)\in D(H(g))$ and $u\in D(H(-g))=D(H(g))$, whence 
$D(H(g))=D(H(g)\P)$.
\par\noindent
(2) From (1) we have $D(\P H(g))= D(H(-g) \P )=D(H(g))$; moreover
$C_0^\infty(\R^d)$ is a core for both operators $\P H(g)$ and $H(-g)\P$.
Therefore it is enough to prove that $\P H(g)u=H(-g)\P u$ $\forall\,u\in 
C_0^\infty(\R^d)$. Indeed, if $u\in 
C_0^\infty(\R^d)$ then $\P u\in 
C_0^\infty(\R^d)$ and
$$
\P H(g)u =\P (-\Delta u+x^2\psi+ig Wu)=-\Delta\P u+
x^2\P u-igW\P u=H(-g)\P u
$$
(3) Again it is enough to prove the identity for $u\in 
C_0^\infty(\R^d)$. By direct inspection:
$$
\overline{\P H(g)\psi}=\overline{H(-g)\P \psi}=\overline{-\Delta\P\psi+
x^2\P\psi-igW\P\psi}=H(g)\P\overline{\psi}
$$
because $\overline{\P\psi}=\P \overline{\psi}$. This proves the 
Proposition.
\vskip 0.2cm\noindent
 {\bf Proof of Theorem 1.1}.
\par\noindent
1. Since $\P$ is continuous in ${\cal H}$ we have (\cite{Ka}, Problem 5.26) 
$Q(g)^\ast =(\P H(g))^\ast =H(g)^\ast\P=\P H(g)$, where the last equality
follows from Assertion (2) of Proposition \ref{prop1}. Since $\P 
H(g)=Q(g)$,
$Q(g)^\ast=Q(g)$. Same argument for
$ Q^\prime (g)$.
\par\noindent
2. The domain of $H(g)$ does not depend on $g$ by (N2) for $g\in{\cal 
S}_1\cup {\cal
S}_2$. Hence also the domain of $Q(g)$ is $g-$independent. Moreover the
scalar products $\la Q(g)u,u\ra$ are obviously entire holomorphic
functions of $g$ $\forall\,u\in D(Q(g))$. Thus $Q(g)$ is by definition a
type-A holomorphic family in the sense of \cite{Ka} (Section VII.1.3) in 
the stated
domain.  We now verify that
$\rho(Q(g))\neq
\emptyset$ for $g$ belonging to a neighbourhood of $\R_+$.  Since $0\notin
\sigma(H_0)$, by (\ref{unif}) with $\t=0$ there is $B>0$ such $H(g)^{-1}$ 
is
uniformly bounded in $\tilde{S}:=\{g\in \S_1\cup\S_2, |g|<B\}$. Hence 
$\mu=0$
is not an  eigenvalue of $Q(g)=\P H(g)$ because $\P$ is invertible.
Therefore $Q(g)$ is invertible for $g\in \tilde{S}$. Now $Ran(Q(g)=L^2$: if
indeed $v\in L^2$, then $\P v\in R(H(g))=L^2$, i.e. there exists $u\in
D(H(g))$ such that $H(g)u=\P v$. Hence $\P H(g))u=v$ and $v\in Ran(\P
H(g))$.  The inverse $Q(g)^{-1}=H(g)^{-1}\P$ is compact as the product of
the compact operator $H(g)^{-1}$ times the continuous operator $\P$. 
Since $Q(g)$ is self-adjoint for $g\in\R$, the compactness of the resolvent
$[Q(g)-z]^{-1}$ extends to all $g$ in a neighbourhod of the real axis (see
\cite{Ka}, Thm VII.2.8).
\par\noindent
3.  Let us first prove the coincidence between the eigenvalues of $Q(g)$
and those of $Q^\prime(g)$. We have:
$$
\P H(g)\psi=\l\psi\Longleftrightarrow H(g)\psi=\l \P \psi
\Longleftrightarrow (H(g)\P)\P\psi=Q^\prime(g)\P\psi=\l \P \psi
$$
Hence $\l$ is eigenvalue of $Q(g)$ with eigenvector $\psi$ if and only if 
$\l$ is eigenvalue of $Q^\prime (g)$ with eigenvector $\P\psi$. 
\newline
  Let now $\mu$ be any
eigenvalue of $Q=\P H$, and let $\psi$ be any corresponding eigenvector.
Then, by the self-adjointness of $\P H$:
$$
Q\psi=\P H\psi=\mu\psi \Longrightarrow  H^\ast H\psi=H^\ast\P \P
H\psi=Q^2\psi=\mu^2\psi.
$$
Thus $\mu^2$ is an eigenvalue of $ H^\ast H$ with the same eigenvector of
$Q$. On the other hand, since, as we have seen, $Q^{-1}$ exists
and is compact the eigenvectors of $Q$ form a complete set. Therefore 
$\mu^2$ is an eigenvalue of
$H^\ast H$ if and only if $ \mu$ or $-\mu$ is an eigenvalue of
$Q$.  
 This completes the proof of
Theorem
\ref{teor1}.
\vskip 0.2cm\noindent
{\bf Proof of Corollary \ref{can}}.
\par\noindent
By the spectral theorem we have, if $u\in D(Q)$:
$$
Qu=\P H u=\sum_{n=0}^\infty \mu_n\langle u,\psi_n\rangle\psi_n
$$
(counting multiplicities). Since $\P Q=H$, and $\P$ is continuous:
$$
H u=\sum_{n=0}^\infty \mu_n\langle u,\psi_n\rangle\P\psi_n, \quad
\forall\,u\in D(H(g))
$$
Now $(\P\psi_n)(x)=\psi_n(-x)$
\vskip 0.2cm\noindent
Define now $Q(\b,\t):=\P H(\b,\t)$ and let us prove that this operator
family enjoys the same properties of $H(\b,\t)$. We have:
\begin{proposition}
\label{propo2}
$Q(\b,\t)$ defined on $D(Q(\b))=D(H(\b))$ is a type-A holomorpic family
with compact resolvents in a neighbourhood of $\R_+$ for $\b$ and $\t$
such that
$(s,t)\in{\cal  R}$,
$s=\arg\b$, $t=\Im\t$. Moreover $C_0^\infty(\R^d)$ is a core of $Q(\b,\t)$. 
Analogous results hold for the \op\ family $Q(\b,\t)$ for $\b$ and $\t$ 
such
that $(s,t)\in{\cal  R}^\prime$, and
$Q(\b,\t)^\ast=Q(\overline\b,\overline\t)$. 
\end{proposition}
{\bf Proof}.
\par\noindent
 The fact that $Q(\b,\t)$ is closed on $D(H(\b))=D(H(\b,\t))$
can be proved by the same argument of Proposition 2.1, (1). To complete
the proof we then proceed as in Theorem 1.1, (2). This proves the
proposition.
\vskip 0.3cm\noindent
{\bf Proof of Theorem 1.3}
\par\noindent
Set $T(\b,\t):=e^{2\t}Q(\b,\t)=\P K(\b,\t)$. Given the analyticity property
of the operator family
$Q(\b,\t)$, we have only to verify the analogous of N5); namely that, for
all
$(\b,\t)$ such that
$(s,t)\in {\cal R}$, the following two properties hold:
\begin{itemize}
\item[(i)] If $\l\notin\sigma(T(0,\t))$, then $\l\in\tilde{\Delta}_1$ 
where:
\begin{eqnarray}
\label{unif1}
\tilde{\Delta}_1:=\{z\in\C:z\notin \sigma(T(\b,\t));
\|[z-(T(\b,\t)]^{-1}\|
\\
\mathrm{is}\; \mathrm{uniformly}\;\mathrm {bounded} 
\;\mathrm {for}\;|\b|\;
\to \; 0 \} ;
\nonumber
\end{eqnarray}
\item[(ii)] If $\l\in\sigma(T(0,\t))$, then $\l$ is stable with respect to
the operator family $T(\b,\t)$. 
\end{itemize}
To prove these assertions, we generalize the argument of \cite{Ca} 
valid for $d=1$. First set
$\rho:=|\b|$, $K(\rho):=K(\b,\t)$, $T(\rho):=T(\b,\t)$.  The proof of
N10) 
 relies on the following results (see \cite{HV}):
\begin{itemize}
\item[(a)] $\ds \lim_{\rho\downarrow 0}K(\rho)u=K(0)u$, 
$\ds \lim_{\rho\downarrow 0}K(\rho)^\ast u=K(0)^\ast u$, $\forall\,u\in
C_0^\infty(\R^d)$;
\item[(b)] $\ds \tilde{\Delta}_1\neq \emptyset$;
\item[(c)] Let $\chi\in\Cinf$ be such that $0\leq \chi(x)\leq 1$,
$\chi(x)=1$ for$|x|\leq 1$, $\chi(x)=0$ for$|x|\geq 2$. For $h\in\N$ let
$\chi_h(x):=\chi(x/h)$, and $M_h(x)=1-\chi_h(x)$. Then:
\begin{itemize}
\item[(1)] If $\r_m\downarrow 0$ and $u_m\in D(K(\r_m))$ are two sequences
such that $\|u_m\|\to 1$, $u_m\to 0$ weakly, and $\|(K(\r_m))u_m\|$
is bounded in $m$, then there exists $a>0$ such that
$$
\liminf_{m\to\infty}\|M_hu_m\|\geq a>0, \quad \forall\,h
$$
\item[(2)] For some $z\in\tilde{\Delta}_1$:
$$
\lim_{h\to\infty}\|[M_h,K(\r)][z-K(\r)]^{-1}\|=0
$$
\item[(3)] $\ds \lim_{h\to\infty\atop \r\downarrow 0}d_h(\l,\r)=+\infty$
$\forall\,\l\in\C$, where:
$$
d_h(\l,\r):=\inf \{\|[\l-K(\r)]M_hu\|:u\in D(K(\r)), \|M_hu\|=1\}
$$
\end{itemize}
\end{itemize}
Hence we must verify the analogous properties, denoted $(a^\prime)$, 
$(b^\prime)$, $(c^\prime)$, for the operator family $T(\r)$. Remark that,
as in \cite{Ca}, the verification of (b') requires an argument completely
independent of \cite{HV} because the operator family $T(\rho)$ is not
sectorial. We have:
\par\noindent
$(a^\prime)$ From $(a)$ and the continuity of $\P$ we can write
$$
\lim_{\r\downarrow 0}T(\r)u=T(0)u,\quad \lim_{\r\downarrow
0}T(\r)^\ast u=T(0)^\ast u,\quad \forall\,u\Cinf
$$
$(b^\prime)$ First remark that $0\in\tilde{\Delta}$ by N9) (i) since
$0\notin \sigma(K(0,\t))$. Then  there is $B>0$ such
that
$$
 \sup_{0\leq |\b|<B}\|K(\b,\t)^{-1}\|<+\infty.
$$
 To prove the analogous
bound with $T(\b,\t)$ in place of $K(\b,\t)$, note that $\P K(\b,\t)\psi=0$
if and only if $\psi=0$. Hence there exists $B>0$ such that $\mu=0$  is not
an eigenvalue of
$T(\b,\t)$ for $|\b|<B$. Thus $T(\b,\t)$ is invertible.  Its range is 
${\cal
H}$: if $v\in{\cal H}$, then $\P v\in Ran(K(\b,\t))={\cal H}$, i.e. there
exists $u\in D(K(\b,\t))$ such that $K(\b,\t)u=\P v$. Thus $\P K(\b,\t)u=v$
and $v\in Ran(\P K(\b,\t))$. Finally $T(\b,\r)^{-1}=(\P K(\b,\t))^{-1}=
K(\b,\t)^{-1}\P$ is uniformly bounded for $|\b|<B$ because $\P$ is bounded
and $K(\b,\t)^{-1}$ is uniformly bounded.
\par\noindent
$(c^\prime)$ Let $\chi\in\Cinf$ be as in $(c)$ with the additional
condition $\chi(x)=\chi(-x)$, i.e. $\P \chi=\chi$. Then $\P \chi_h=\chi_h$,
and
$\P M_h=M_h$. We have:
\par\noindent
\begin{itemize}
\item[(1')] Let $\r_m\downarrow 0$ and $u_m\in D(T(\r_m))$ be such that
$\|u_m\|\to 1$, $u_m\to 0$ weakly and $\|T(\r_m)u_m\|\leq {\rm const.}$
$\forall\,m$. Then $\|K(\r_m)u_m\|=\|T(\r_m)u_m\|\leq {\rm const.}$; hence
by (c1) there exists $a>0$ such that
$$
\liminf_{m\to\infty}\|M_hu_m\|\geq a>0,\quad \forall\,h
$$
\item[(2')] As proved in \cite{HV}, if (c2) holds for some
$z\in\tilde{\Delta}_1$ then it holds for all $z\in\tilde{\Delta}_1$. Thus 
we
can take $z=0\in 
\tilde{\Delta}\cap \tilde{\Delta}_1$ and we have:
\begin{eqnarray*}
\lim_{h\to\infty}\|[M_h,T(\r)](\P K(\r))^{-1}\|=
\lim_{h\to\infty}\|(M_h\P K(\r)-\P K(\r)M_h)(\P K(\r))^{-1}\|
\\
=\lim_{h\to\infty}\|\P [M_h,K(\r)]K(\r))^{-1}\P\|=0\qquad\qquad\qquad\qquad
\end{eqnarray*}
where the last equality follows from the unitarity of $\P$ and (c2). 
\item[(3')] Let $\l\in \C$ and
$$
d^{\prime}_h(\l,\r):=\inf \{\|(\l-T(\r))M_hu\|:u\in D(T(\r)), \|M_hu\|=1\}
$$
Then:
\begin{eqnarray*}
\|[\l-T(\r)]M_hu\|=\|[\l(1-\P)+\P(\l-K(\r))]M_hu\|\geq 
\\
\|[\l-K(\r)]M_hu\|-|\l|\|(1-\P)M_hu\|\geq \|[\l-K(\r)]M_hu\|-|\l|
\end{eqnarray*}
Hence $d^{\prime}_h(\l,\r)\geq d_h(\l,\r)-|\l|$ and by (3) $\ds
\lim_{h\to\infty}d^{\prime}_h(\l,\r)=+\infty$. The assertion is now a
direct application of \cite{HV}, Theorem 5.4. This concludes the proof of
Assertions 1 and 2 of Theorem 1.3.  
\end{itemize}

\vskip 0.3cm\noindent
Let us now turn to the proof of Assertion 3, i.e. the Borel summability of
the eigenvalues of the operator family $Q(g,\t):=Q(\b,\t)$ for $\b=ig$,
$-\pi/4<{\rm arg}g <\pi/4$, $|g|$ suitably small (depending on the
unperturbed eigenvalue).  

To this end, 
we adapt to the present situation the proof \cite{Na} valid for the
\op\ family $H(g,\t):=H(\b,\t), \b=ig$, in turn based on the general
argument of
\cite{HP}. 

First remark that if $(\b,\t)$ generates the parallelogram ${\cal  R}$
defined in (\ref{par}) then $(g,\t)$ generates the parallelogram
\be
\label{par1}
\widehat{\cal  R}=\{(s,t)\in\R^2:-\pi/2<(2K-1)t+s<\pi/2,
-\pi/2<(2K+3)t+s<\pi/2\},
\;
\ee 
where now $s={\rm arg}\,g= {\rm arg}\,\b-\pi/2$. From now on, with abuse of
notation,
we write $(g,\t)\in \widehat{\cal  R}$ whenever $(s,t)\in{\cal R}$. 

Let 
$\l$ be an eigenvalue of $H_0(\t):=H(0,\t)$ of multiplicity $m(\l):=m$. 
Denote
$P(0,\t)$ the corresponding projection. By the above stability result, this
means that if
$\Gamma$ is a circumference of radius
$\epsilon$ centered at $\l$ there is $C>0$ independent of $(g,\t)\in 
\widehat{\cal
R}$ such that, denoting
$R_Q(z,g,\t):=[Q(g,\t)-z]^{-1}$ the resolvent of $Q(g,\t)$:
$$
\sup_{z\in\Gamma_0}\|[Q(g,\t)-z]^{-1}\|\leq C, \quad |g|\to 0
$$
and that  ${\rm dim}\,\widehat{P}(g,\t)={\rm dim}\,\widehat{P}$ as
$|g|\to 0$,
$(g,\t)\in \widehat{\cal R}$, ${\rm arg}\,g$ fixed. This time:
\be
\label{proiettore}
\widehat{P}(g,\t):=\frac{1}{2\pi i}\int_{\Gamma}R_Q(z,g,\t)\,dz, \quad
\widehat{P}\equiv \widehat{P}(0,\t):=
\frac{1}{2\pi i}\int_{\Gamma}R_Q(z,0,\t)\,dz
\ee
are the projections on the parts of $\sigma(Q(g,\t))$,
$\sigma({\P} H(0,\t))$ enclosed in $\Gamma$. We recall that 
$\sigma(Q(g,\t))$ is independent of $\t$ for all $(g,\t)$ in the stated
analyticity region, and that $\widehat{P}(0,\t)=P(0,\t)$.   It follows that
$Q(g,\t)$ has exactly
$m$ eigenvalues (counting multiplicities) in $\Gamma$, denoted once again
$\mu_1(g),\ldots,\mu_m(g)$. We explicitly note that, unlike the $m=1$ case,
when the unperturbed eigenvalue is degenerate, the analyticity of the \op\
family does not a priori entail the same property of the eigenvalues
$\mu_1(g),\ldots,\mu_m(g)$, so that the analysis of \cite{Na},\cite{HP} is
necessary.  Following [\cite{HP}, Sect.5] set:
$$
{\cal M}(g,\t):=Ran(\widehat{P}_Q(g,\t)); \qquad \widehat{D}(g,\t)
:=\widehat{P}(0,\t)\widehat{P}(g,\t)\widehat{P}(0,\t)
$$
Under the present conditions $\widehat{D}(g,\t)$ is invertible on ${\cal
M}(0):=Ran(\widehat{P}(0,\t))$. Hence the present problem can be reduced to 
a
finite-dimensional one in
${\cal M}(0,\t)$ by setting
\begin{eqnarray*}
E(g,\t)&:=&\widehat{D}(g,\t)^{-1/2}N(g,\t)\widehat{D}(g,\t)^{-1/2}; \\
N(g,\t)&:=&\widehat{P}(0,\t)\widehat{P}(g,\t)[Q(g,\t)-\l]\widehat{P}(g,\t)
\widehat{P}(0
,\t)
\end{eqnarray*}

\par\noindent
As in [\cite{HP}, Thms 4.1, 4.2]  the \RS\ series for each
eigenvalue
$\mu_s(g): s=1,\ldots,m$ near $\l$ is Borel summable upon verification of
the two following assertions: there exist $\eta(\delta)>0$ and a sequence 
of linear
\op s $\{E_i(0,\t)\}$ in ${\cal M}(0,\t)$ such that
\begin{itemize}
\item[(i)] $E(g,\t)$ is an \op -valued analytic function for
$(g,\t)\in\widehat{\cal R}$; As we know, this entails that $E(g)$ 
is is an \op -valued analytic function in the sector
$$
{\cal S}_{K,\delta}:=\{g\in\C: 0<|g|<\eta(\delta);
-(2K-1)\frac{\pi}{2}+\delta<{\rm arg}\,(g)<(2K+3)\frac{\pi}{2}-\delta\}
$$
\item[(ii)] $E(g,\t)$ fulfills a strong asymptotic condition  in
$\widehat{\cal R}$ (and thus, in particular, for $g\in{\cal
S}_{K,\delta}$) and admits
$\ds
\sum_{i=0}^\infty E_i(0,\t)g^i$ as asymptotic series; namely, there exist
$A(\delta)>0$,
$C(\delta)>0$ such that
\be
\label{Sac}
||R_N(g)\|:=\|E(g,\t)-\sum_{i=0}^{N-1} E_i(0,\t)g^i\|\leq
AC^N\Gamma((2K-1)N/2)|g|^N
\ee
as $|g|\to 0$, $(g,\t)\in \widehat{\cal R}$, $g\in {\cal S}_{K,\delta}$;
\item[(iii)] $\qquad E_i(0,\t)=E^\ast_i(0,\t)$, $\quad i=0,1,\ldots$,
$\quad\t\in\R$. 
\end{itemize} 
Given the stability result (Assertion 2 of the present Theorem 1.3) the 
proof of (i)
and (iii) is
identical to that of \cite{Na}, Lemma 2.5 (i) and is therefore omitted.  We
prove  assertion (ii).  Under the present conditions 
the Rayleigh-\Sc\
perturbation expansion is generated by inserting in (\ref{proiettore}) the
 (formal) expansion of the resolvent $R_Q(z,g,\t):=[Q(g,\t)-z]^{-1}$:
\be
\label{Neumann}
R_Q(z,g,\t)=R_Q(z,g,\t)\sum_{p=0}^{N-1}[igWR_\P(z,0,\t)]^p+R_Q(z)[igWR_\P(z
,0,\t)]^N
\ee
and performing the contour integration. Moreover (see once more \cite{HP},
Section 5.7), to prove (\ref{Sac}) it is enough to prove the analogous 
bound on
$\widehat{D}(g,\t)$ and
$N(g,\t)$.  Since
$\widehat{D}(g)=
\widehat{{P}}(0,\t)\widehat{{P}}_Q(g,\t)\widehat{P}(0,t)$, we have,
inserting  (\ref{Neumann}) 
\begin{eqnarray*}
D_N(g,\t)&:=&D(g,\t)-\sum_{i=0}^{N-1}
D_i(0,\t)g^i
\\
&=&\widehat{P}(0,\t)\frac{1}{2\pi i}
\int_{\Gamma_0}R_Q(z,g,\t)[W(x)R_\P(z,0,\t)]^N\widehat{{P}}(0,\t)
\end{eqnarray*}
By the analyticity and uniform boundedness of the
resolvent  $R_Q(z,g,\t)$  in $\widehat{\cal R}$ (and hence in particular 
for
$g\in\S_{K,\delta}$),  it is enough to prove the estimate
\be
\label{stima1}
\sup_{z\in\Gamma_j}\|[igWR_\P(z,0,\t)]^N \widehat{P}_0\|\leq
AC^N\Gamma((2K-1)N/2)|g|^N
\ee
In turn, since $\widehat{P}(0,\t)= P(0,\t)$, by the Combes-Thomas argument
(see
\cite{HP}, Sect. 5 for details) to prove  (\ref{stima1}) it is enough to to
find a function
$f:\R^d\to\R$ such that 
\be
\label{stima2}
\|e^fP(0,\t)\|<+\infty;\qquad \sup_{x\in\R^d}|W(x)e^{-f/N}|\leq 
N^{\frac{2K-1}{2}}
\ee
Now a basis in $Ran(P_j)$ is given by $m$ functions of the type
$$ {\cal Q}(e^{\t/2}x_1,\ldots,e^{\t/2}x_d)e^{-e^{\t/2}|x|^2}
$$ 
where
${\cal Q}$ is a polynomial of degree at most
$m$. Therefore both estimates are fulfilled by choosing $\ds f=\alpha
|x|^2$ with
$\alpha=\alpha(\t) <1/2$. This condition is always satsfied if $(g,\t)\in
\widehat{\cal R}$ because $|{\rm Im}\,\t|<\pi/4$.  This concludes the proof 
of
the Theorem.
 \vskip 0.3cm\noindent
{\bf Remark}
\par\noindent
The summability statement just proved, called Borel summability for the 
sake of
simplicity, is more precisely the Borel-Leroy summability of order
$q:=(K-1)/2$. 
\section{Conclusion}
Even though the object of main physical interest are the eigenvalues of 
$H(g)$ rather 
than its singular values $\mu_k(g)$ determined in this paper, the singular 
values
yield a property that the eigenvalues cannot in general yield since the 
operator
$H(g)$ is not normal: namely, a diagonal form. If an operator is physically
interesting a diagonalization of it  is clearly useful. 
To examine this point in more detail,  consider 
once again the canonical expansion (\ref{canonical}) of Corollary 1.2:
$$
\label{canonexp}
H(g)u=\sum_{k=0}^{\infty}\mu_k(g)\langle u,\psi_k\rangle\P\psi_k, \quad 
u\in
D(H(g))
$$
Since both vector sequences $\{\psi_k\}$  and $\{\P\psi_k\}$ are 
orthonormal 
we have
\be
\label{dd}
\langle \P\psi_k,H(g)\psi_l\rangle=\mu_k(g)\delta_{k,l}
\ee
Moreover the orthonormal sequences $\{\psi_k\}$  and $\{\P\psi_k\}$ are 
complete in
the Hilbert space. Hence formula (\ref{dd}) is an actual diagonalization of 
$H(g)$.
The basis $\{\psi_k\}$  acts in the domain, and the basis $\{\P\psi_k\}$ in 
the
range.
A complete diagonalization of the $\PT$-symmetric but non-normal operator 
$H(g)$ has
been therefore obtained: the singular values $\mu_k(g)$ and the  
eigenvectors
$\psi_k$ (and thus also the vectors $\P\psi_k$) are indeed
uniquely defined  by  perturbation theory  through the Borel summability.

More precisely, the general formula (\ref{cc1})
$$
Hu=\sum_{k=0}^{\infty}\mu_k\langle u,\psi_k\rangle\psi^\prime_k, \quad
u\in D(H)
$$
which provides a diagonalization for an operator $H$ with compact resolvent 
with respect 
to the pair of orthonormal bases $\{\psi_k\}$ and $\{\psi^\prime_k\}$,
requires a priori the computation of  $\mu_k$ and $\psi_k$ as solutions
of the spectral problem
\begin{equation}
\label{ccc}
H^\ast(g) H(g)\psi=\mu^2\psi
\end{equation}
which represents an eigenvalue problem more complicated than 
$H(g)\phi=\lambda\phi$.   
The result of this paper means that the eigenvalue 
problem (\ref{ccc}) can be replaced by the more tractable one
$$
H(g)\psi=\mu\P\psi
$$
which can be solved by perturbation theory and Borel summability.
\vfill\eject
\vskip 0.3cm\noindent
 {\bf Acknowledgment}
\par\noindent
We thank Francesco Cannata for his interest in this work and several
useful suggestions.

\end{document}